\documentclass[conference]{IEEEtran}
\IEEEoverridecommandlockouts
\usepackage{cite}
\usepackage{amsmath,amssymb,amsfonts}
\usepackage{algorithmic}
\usepackage{graphicx}
\usepackage{textcomp}
\usepackage{xcolor}
\usepackage{siunitx}
\usepackage{url}
\usepackage{mathptmx} 

\usepackage{float}
\usepackage[caption=false]{subfig}

\DeclareUnicodeCharacter{FF0C}{~}

\usepackage[numbers,sort&compress]{natbib}

\def\BibTeX{{\rm B\kern-.05em{\sc i\kern-.025em b}\kern-.08em
    T\kern-.1667em\lower.7ex\hbox{E}\kern-.125emX}}

\usepackage{threeparttable}
\usepackage{array}

\usepackage[colorinlistoftodos,prependcaption,textsize=tiny]{todonotes}
 
\begin{document}

\bstctlcite{IEEEexample:BSTcontrol}

\title{Fast Switching Serial and Parallel Paradigms of SNN Inference on Multi-core Heterogeneous Neuromorphic Platform SpiNNaker2\\
}


\author{
    \IEEEauthorblockN{Jiaxin Huang\IEEEauthorrefmark{1}\IEEEauthorrefmark{2}, 
    Bernhard Vogginger\IEEEauthorrefmark{2},
    Florian Kelber\IEEEauthorrefmark{2},
    Hector Gonzalez\IEEEauthorrefmark{1}\IEEEauthorrefmark{2},
    Klaus Knobloch\IEEEauthorrefmark{4}, 
    Christian Georg Mayr\IEEEauthorrefmark{2}\IEEEauthorrefmark{3}}
    \IEEEauthorblockA{\IEEEauthorrefmark{1}SpiNNcloud Systems
    \\\{jiaxin.huang, hector.gonzalez\}@spinncloud.com}
    \IEEEauthorblockA{\IEEEauthorrefmark{2}Technische Universität Dresden
    \\\{Florian.Kelber, Bernhard.Vogginger, christian.mayr\}@tu-dresden.de}
    \IEEEauthorblockA{\IEEEauthorrefmark{3}Centre for Tactile Internet (CeTI) with Human-in-the-Loop, Cluster of Excellence, Technische Universität Dresden}
    \IEEEauthorblockA{\IEEEauthorrefmark{4}Infineon Technologies Dresden
    \\\{Klaus.Knobloch\}@infineon.com}
}

\maketitle

\begin{abstract}
With serial and parallel processors introduced into Spiking Neural Networks (SNNs) execution, more and more researchers are dedicated to improving the performance of the computing paradigms by taking full advantage of the strengths of the available processor. In this paper, we compare and integrate serial and parallel paradigms into one SNN compiling system. For a faster switching between them in the layer granularity, we train the classifier to prejudge a better paradigm before compiling instead of making the decision afterward, saving a great amount of compiling time and RAM space on the host PC. The classifier Adaptive Boost, with the highest accuracy (91.69\%) among 12 classifiers, is integrated into the switching system, which utilizes less memory and processors on the multi-core neuromorphic hardware backend SpiNNaker2 than two individual paradigms. To the best of our knowledge, it is the first fast-switching compiling system for SNN simulation.


\end{abstract}


\section{Introduction}\label{sec_introduction}
By mimicking the information-processing activity of biological brains, SNNs are expected to be more powerful and energy-efficient than conventional neural networks. A wide variety of dedicated neuromorphic hardware, such as Loihi\cite{loihi}, BrainScaleS\cite{brainscales_ref1}\cite{brainscales_ref2}, and SpiNNaker\cite{spi1_ref}, have been designed to perform SNN inference, simulating the process of synaptic processing and neural update. Some neuromorphic simulators \cite{francisco_1}\cite{GENN_for_snn} use off-the-shelf CPU or GPU as the hardware backends. SpNNaker~2 is a neuromorphic platform integrating multi-core distributed serial and parallel processors. Using the ARM processor, the serial SNN inference paradigm on SpiNNaker2 fully utilizes the input sparsity to achieve energy savings. This paradigm's event-based mechanism enables the serial processor to update the neural dynamics only when perceiving the connected pre-neuron fires. However, running this paradigm requires a relatively complex data structure, which could be very large for dense synaptic connections.
The parallel paradigm, reported in \cite{ref_jennifer_aicas23}\cite{ref_jennifer_frontiers}, is intended to accelerate the conventional serial processing paradigm by activating the parallel processor MAC array by SNN execution.
Although papers \cite{ref_jennifer_aicas23}\cite{ref_jennifer_frontiers} have deployed a series of optimization strategies to alleviate the memory weakness derived from operands' zero padding and potential sparse synaptic connection, the optimization effect is not always apparent in various situations. The question of which paradigm is better regarding memory usage for various SNN layers is one of the subjects of this paper.

According to the analysis above, it is known that the serial and parallel paradigms have strengths and weaknesses in memory performance that balance each other out. By integrating them into a comprehensive compiling system, the SNN inference execution system can adapt to different scenarios and obtain a better spatial performance. 

As for the related studies, EDLUT \cite{francisco_1}\cite{francisco_2}, as a spiking neural simulator, has implemented two different neural dynamic evaluation techniques. The performances of these two techniques are compared when simulating the cerebellum's microzone(s). The researchers conclude that the two techniques outperform each other for small- and large-size neurons, respectively. However, they did not investigate the specific switching conditions of the two techniques, and their compiling system cannot provide a quick decision on which technique to choose when getting a new model of moderate size of neurons. Each new SNN model needs to be realistically compiled before knowing which technique performs better.

\begin{figure*}[h]
\centering
\includegraphics[width=0.9\linewidth]{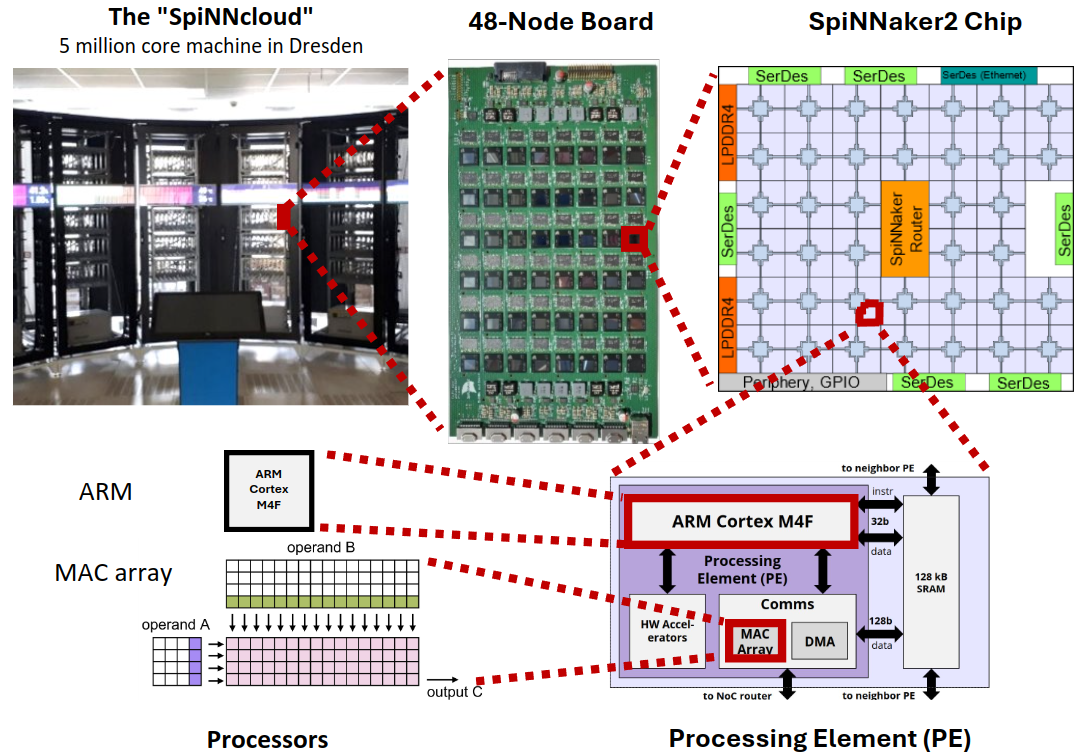}
\caption{Overview of SpiNNaker2 architecture.}
\label{spi2_hw}
\end{figure*}

This paper abstracts the problem of paradigm/technique selection as a classification issue. We train 12 classifiers and deploy the one with the best performance as the prediction tool to enable fast switching of serial and parallel paradigms before compiling to achieve a better memory performance when executing SNN inference on the multi-core heterogeneous neuromorphic platform SpiNNaker2. To the best of our knowledge, this is the first work to prejudge SNN execution paradigms before compiling to achieve an efficient SNN deployment and simulation.

In this paper, we briefly introduce the hardware backend SpiNNaker2 in Section \ref{sec_spi2} before elaborating on the serial and parallel paradigms in Section \ref{sec_ser_par_paradigms}. Next, we analyze and evaluate the switching system in Section \ref{sec_comp_and_fus}. Finally, Section \ref{sec_conclusion} makes a conclusion.

\section{Hardware backend: SpiNNaker2}\label{sec_spi2}
SpiNNaker2 \cite{Mayr_ref} is a massively parallel computing system that can be scaled from one chip with 152 cores \cite{Amir} to a supercomputer scale with millions of cores. Each core (a processing element (PE)) has one ARM Cortex M4F and one MAC array. The instructions and data compiled with serial or parallel paradigm that will be described in Section \ref{sec_ser_par_paradigms} are loaded to 128kB SRAM of each PE. All PEs communicate with each other by leveraging the Network-on-Chip (NoC) architecture.

Since the serial and parallel paradigms are highly related to the processors involved in PE, and the ARM processor is well-known, we focus on introducing the dedicated MAC array on SpiNNaker2.
The MAC array on one PE has 64 MAC units in a $4\times16$ layout, as stated in \cite{Yexin_new_ref}\cite{Zeinolabedin_ref}. Executing matrix multiplication requires operand memory alignment to adapt to this hardware architecture. The precision of operands could be 8-bit or 16-bit, and the output precision can be configured to 8-/16-/32-bit.

\section{Serial and parallel paradigms}\label{sec_ser_par_paradigms}
In this section, we elaborate on the serial and parallel paradigms in terms of mapping and execution based on SpiNNaker2's multi-core heterogeneous processor hardware architecture.

\begin{figure*}[htbp]
\centering
\includegraphics[width=1\linewidth]{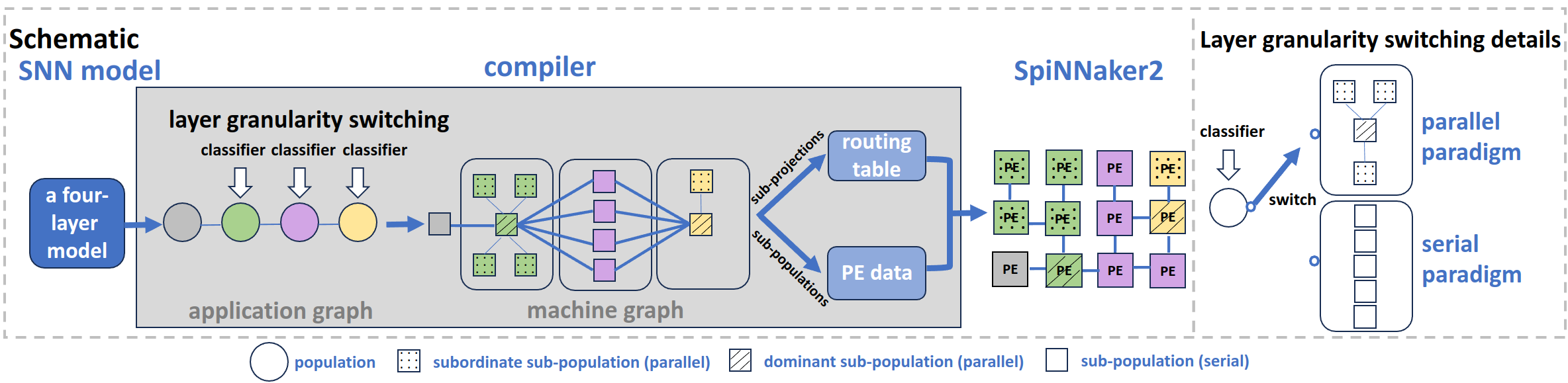}
\caption{Schematic of mapping the SNN model on SpiNNaker2 with the switching system, and the details of layer granularity switching.}
\label{pic:big_diagram}
\end{figure*}


In general, mapping the trained SNN model to neuromorphic hardware for inference execution requires a series of transformation steps, such as parsing and preprocessing SNN model information, as Figure \ref{pic:big_diagram} demonstrates. The transformation steps start from the specifically trained or ANN-converted SNN model. The SNN model is interpreted into an application graph, a concept from \cite{spynnaker_ref}. Usually, each vertex of the application graph contains all neurons of one layer, and edges indicate the projections of the inter- and inner-layer. The neuron population in each vertex is then split into one or several sub-populations to fit the SRAM resource of each PE. All the sub-populations and the corresponding projections between them form a machine graph. The connection relations of these sub-populations contribute to generating a routing table. Finally, this transformed information is 
loaded on SpiNNaker2 before execution. This mapping framework enables large-scale SNN simulation on multiple PEs of SpiNNaker2. 

Under this mapping framework, we detail two paradigms according to the processors deployed in the SNN inference execution on SpiNNaker2. The serial paradigm of utilizing the ARM processor follows the model partitioning approach from SpyNNaker \cite{spynnaker_ref}, splitting the application graph based on the pre-defined neuron number capacity of each PE (255). The parallel paradigm proposed in \cite{ref_jennifer_aicas23}\cite{ref_jennifer_frontiers} introduces the MAC array, a machine learning accelerator on SpiNNaker2, into the SNN hardware simulation process. Unlike the serial paradigm, the parallel paradigm considers the dynamic adjustment of neurons and synapses per PE while partitioning the application graph into the machine graph.


\subsection{Serial paradigm}\label{subsec_serial_paradigm}
For the serial paradigm, the event-based synaptic processing and time-triggered neural update are applied to SNN execution during runtime. The event of a spike arrival triggers the synaptic processing in the current PE. Specifically, the source neuron index embedded in the spiking package unlocks an entry of the pre-loaded master population table. This entry points at one item of the address list, indicating the first address and matrix row length of a block of synaptic matrix on local SRAM. Each row within one block saves the synaptic information between the spiked source neuron and one of the target neurons, including weight, delay, synapse type (excitatory or inhibitory), and target neuron index. One source neuron corresponds to one block, and multiple blocks compose the whole synaptic matrix. We accumulate the weights activated by all the spikes arriving at the current PE in the last timestep. Then, we classify them into slots of synaptic input buffer according to delay and synapse type before subtracting 1 for the delay labeled on each slot (0 returns to the largest delay). The weight difference of the two synapse types in slot 0 is regarded as the input current. In the neural update stage, we add the input current of each target neuron with decayed membrane potentials and compare the result with a threshold to decide the neuron status (spike or not). The whole process can be formulated by the following leaky integrate-and-fire (LIF) formula referring to \cite{lif_model}:

\begin{equation}
V_{j}^{t+1} = \Sigma_{i} W_{ji} x_{i}^{t-d(j,i)} + \alpha V_{j}^{t} - z_{j}^{t}V_{th}  \label{eq0}    
\end{equation}

The contents in data structures (master population table, address list, synaptic matrix) and the size of the memory placeholders (input spike buffer, synaptic input buffer) involved in runtime are created by the compiler and loaded to SpiNNaker2 before inference execution. If the number of neurons in the population is larger than 255, the population is equally partitioned into several sub-populations corresponding to the same number of PEs. The data structures and memory placeholders are also split and distributed into these PEs.


\subsection{Parallel paradigm}\label{subsec_parallel_paradigm}
Subsection \ref{subsec_serial_paradigm} mentions that the SNN inference consists of synaptic processing and neural update. The MAC array can accelerate the former \cite{ref_jennifer_aicas23}\cite{ref_jennifer_frontiers}. In this paradigm, the reversed order and input merging table are saved in the dominant PE to pre-process the spikes in the stacked input buffer to adapt to the data layout of the optimized weight-delay-map. Then, the pre-processed spike train is read by subordinate PEs, where the MAC array operates matrix multiplication for the obtained spike train and the optimized weight-delay-map.

The compiler generates the content of data structures (reversed order, input merging table, and optimized weight-delay-map) and the size of the memory placeholders (input spike buffer, stacked input buffer) before loading to neuromorphic hardware. If one subordinate PE does not have sufficient DTCM (data tightly coupled memory) to save the whole optimized weight-delay-map, it will be split into multiple cores in a spatial-temporal balancing way by the two-stage splitting algorithm. Other data structures and memory placeholders remain unchanged.




Not limited by the fixed number of neurons per PE as in the serial paradigm, the parallel considers the impact of both neuron number and weight sparsity on SRAM memory consumption during compiling. It is more friendly to the SNN layer with very sparse connections between large numbers of neurons, and the dense layer with small neuron numbers.

\begin{table*}[htbp]

\caption{cost model in dtcm}
\begin{center}
\begin{threeparttable}
\resizebox{16cm}{!}{
\begin{tabular}{|c|c|c|}
\hline
 &\textbf{\textit{item}}&\textbf{\textit{cost model (Byte)}}\\
\hline
\textbf{\textit{serial}}&input spike buffer & (32/8)*n\_neuron\\
\cline{2-3}
\textbf{\textit{paradigm}} & DMA buffer & 0 (DRAM not involved) \\
\cline{2-3}
& master population table & (96/8)*n\_source\_vertex  \\
\cline{2-3}
& address list & (32/8)*n\_address\_list\_rows  \\
\cline{2-3}
& synaptic matrix & (32/8)*n\_neuron*n\_neuron*max\_connected\_rate  \\
\cline{2-3}
& synaptic input buffer & (16/8)*n\_neuron*delay\_range*n\_projection\_type  \\
\cline{2-3}
& neuron and synapse model & (32/8)*n\_param(LIF:8+6)  \\
\cline{2-3}
& output recording & (32/8)*(ceil(n\_neuron/32)+1)+(32/8)*n\_neuron*3 \\
\cline{2-3}
& stack \& heap & (96/8)*n\_source\_vertex \\
\cline{2-3}
& hw mgmt \& OS & 6000 \\
\hline

\textbf{\textit{parallel}}&input spike buffer & (32/8)*n\_source\_neuron\\
\cline{2-3}
\textbf{\textit{paradigm}} &reversed order & (32/16)*n\_source\_neuron*delay\_range \\
\cline{2-3}
\textbf{\textit{(dominant))}} & input merging table & n\_source\_neuron*delay\_range*3  \\
\cline{2-3}
& stacked input & n\_source\_neuron*delay\_range*4  \\
\cline{2-3}
& neuron and synapse model & (32/8)*n\_neuron*n\_neuron*max\_connected\_rate  \\
\cline{2-3}
& output recording & (32/8)*n\_target\_neuron*4 \\
\cline{2-3}
& stack \& heap & (96/8)*n\_source\_vertex \\
\cline{2-3}
& hw mgmt \& OS & 6000 \\
\hline

\textbf{\textit{parallel}}&optimized weight delay map & (can't be accurately estimated)\\
\cline{2-3}
\textbf{\textit{paradigm}}& output recording & (16/8)*n\_neuron*delay\_range*n\_projection\_type  \\
\cline{2-3}
\textbf{\textit{(subordinate))}}& stack \& heap & (96/8)*n\_source\_vertex \\
\cline{2-3}
& hw mgmt \& OS & 6000 \\
\hline
\end{tabular}
}
\end{threeparttable}
\label{tab1}
\end{center}
\end{table*}

\section{Comparison and integration}\label{sec_comp_and_fus}
As the counterpart and opponent, the serial and parallel paradigms elaborated in Section \ref{sec_ser_par_paradigms} have their own specialization. They can address a broader range of problems than either could on its own when we integrate them into one system, where we can switch them according to different situations with the purpose of less memory cost. The most straightforward approach for paradigm comparison is compiling the given SNN layer and selecting the optimal solution. However, the compiling time and the RAM occupation on the host PC are not negligible, especially for large models (such as the 8 hours required for the microcircuit reported in \cite{microcircuit2018}). The problem of compiling time gets even worse when compiling with two paradigms sequentially. Moreover, saving two compiling results may cause a RAM crisis on the host PC. Thus, we first analyze how various factors of one layer (one population of application graph) of SNN affect the spatial performance difference of two paradigms in Subsection \ref{subsec_comp_analysis}, where the analysis result shows the complexity of this problem. So, in Subsection \ref{subsec_integration}, we abstract this problem, model it to a binary classification issue, and select the best classifier from 12 candidates to solve this problem. Subsection \ref{subsec_evaluation} evaluates the switching system embedded with this fast decision tool.

\begin{figure*}[htbp]
\centering
\includegraphics[width=1\linewidth]{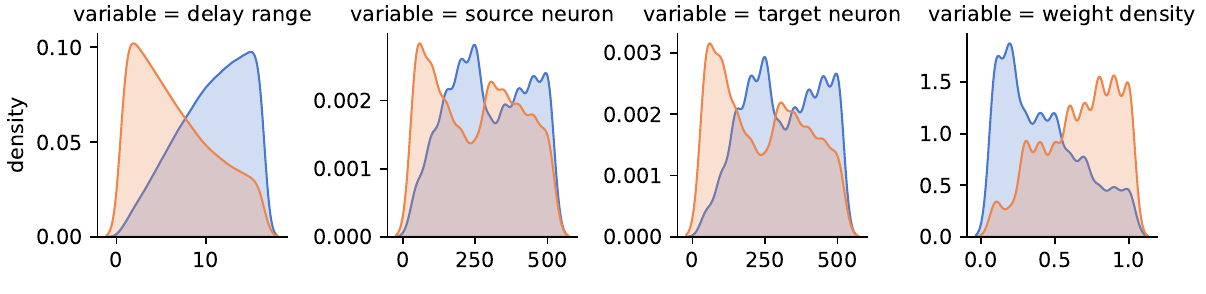}
\caption{The marginal distribution of four univariables based on the acquired dataset. The orange line represents the parallel paradigma, and blue line for serial paradigma.}
\label{pic:univariable_marginal_distribution}
\end{figure*}

\begin{figure*}[htbp]
\centering
\includegraphics[width=1\linewidth]{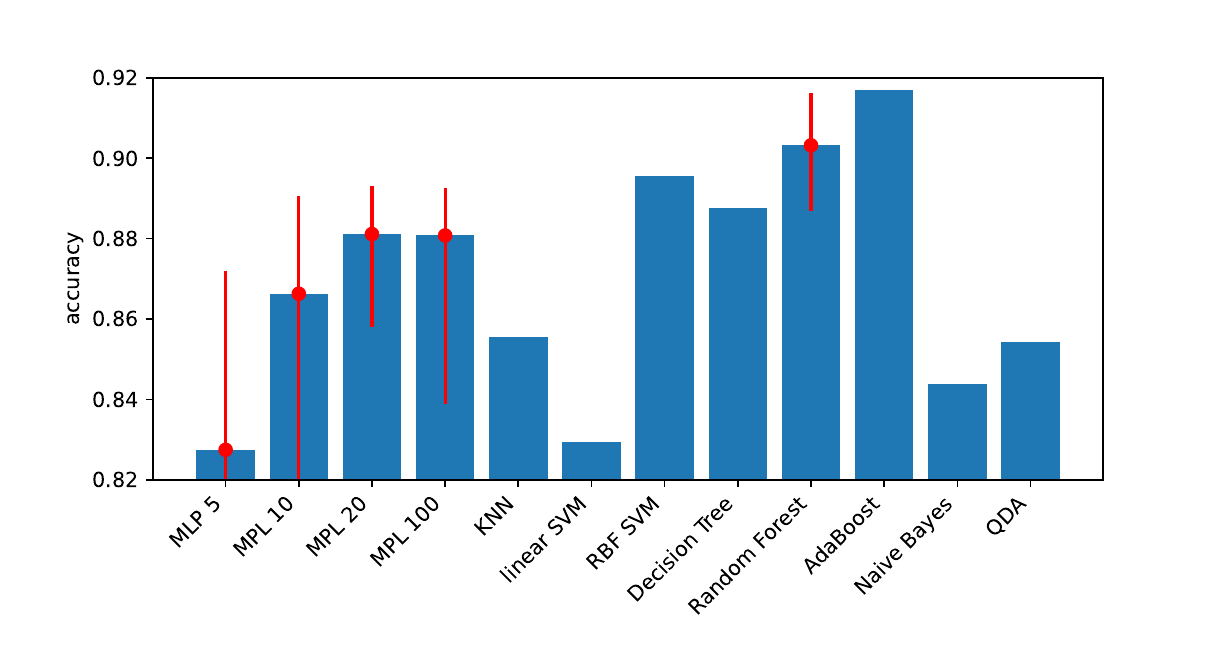}
\caption{Accuracy comparison among 12 classifiers. MLP x means the multilayer perceptron model with x neurons in hidden layer. The red lines mark the accuracy range of training with 20 different random seeds.}
\label{pic:accuracy_comparison}
\end{figure*}

\subsection{Dataset acquisition and statistical analysis}\label{subsec_comp_analysis}
The PE occupation of serial and parallel paradigms is closely bound up with the characters of SNN layers. To investigate the relations of layer characters (delay range, source neuron number, target neuron number, weight density) and the best paradigm (serial or parallel) that requires less PE, we need to collect a large dataset as raw materials for analysis. 

We can calculate the number of PEs for one layer using the serial paradigm. Table \ref{tab1} lists all the data structures in DTCM and their memory cost models. The source and target neuron numbers are fixed to 255 according to \cite{spynnaker_ref}, and we use 8-bit weights. Unlike the implementation of the paper \cite{spynnaker_ref}, we increase the DTCM from 64 kB to 96 kB, considering a larger SRAM space of each PE on SpiNNaker2 than on SpiNNaker. Another difference lies in the exclusion of DRAM in the experiments of this paper, so all the synaptic rows are saved in local SRAM, and the memory placeholder for receiving synaptic rows loaded by DMA from DRAM during runtime is removed. We find that the synaptic matrix, which is proportional to the weight density, dominates the memory summation of data structures in Table \ref{tab1}, and the DTCM of one PE is incapable of holding all the data structures when the weight density is over 25\%. Thus, we equally distribute the synaptic matrix into 2-4 adjacent PEs for the layer with dense weight for the layer with dense weight. We also equally split the source and target neurons when they exceed the 255 limitation.

For the parallel paradigm, it is hard to calculate the required number of PEs. Even if a complex probabilistic model is built to model the optimization rate of four optimization strategies from \cite{ref_jennifer_frontiers} sequentially, the size of the optimized weight-delay-map is still a range instead of an exact value. To obtain the accurate subordinate PE number, we run on parallel paradigm's compiler the randomly generated 16000 SNN layers, whose source and target neurons range from 50 to 500 with step length 50, weight density 10\% - 100\% with 10\% step length, delay range 1 - 16 with step length 1. Within the scope of these settings, one dominant PE is enough according to our calculation based on the cost model in Table \ref{tab1}. The subordinate PE plus the dominant is the total PE quantity.


Fig. \ref{pic:univariable_marginal_distribution} plots the marginal distribution of four factors of SNN layer character, presenting the influence of the univariable on the classification result. This figure shows that the parallel paradigm improves with decreasing delay range and increasing weight density. Nevertheless, the parallel paradigm is not the only winner of all the cases even for a petite delay range and huge weight density. The situation for source and target neurons is even more complex. Therefore, we introduce a classifier as a binary classification tool to solve this problem.


\subsection{Classifier comparison and selection}\label{subsec_integration}
As mentioned at the end of Subsection \ref{subsec_comp_analysis}, the correspondence of input data (four factors of layer character) and labels (switch to which paradigm) is unclear. The classifier is an effective tool for learning features of input data and establishing a high-accuracy input-output correspondence. We train 12 kinds of classifiers with the dataset acquired in Subsection \ref{subsec_comp_analysis}, and the highest accuracy of 91.69\% comes from the Adaptive Boost algorithm, as demonstrated in Fig.\ref{pic:accuracy_comparison}. By integrating this algorithm into the switching system, we can achieve a high-precision fast switching.




\subsection{Evaluation of the classifier integrated switching system}\label{subsec_evaluation}

\begin{figure}[htbp]
\centering
\includegraphics[width=1\linewidth]{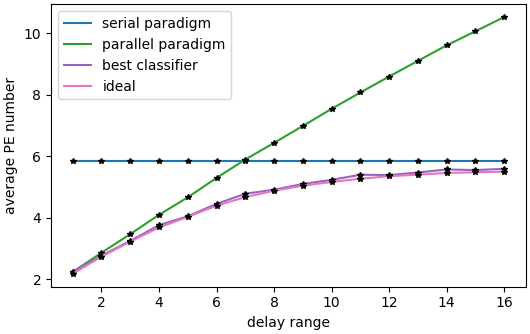}
\caption{Memory performance comparison among two paradigms, trained classifier (Adaptive Boost, estimate before compiling both paradigms), and the ideal situation (classify after compiling both paradigms).}
\label{pic:classifier_evaluation}
\end{figure}

In order to more intuitively evaluate the memory performance of the Adaptive Boost algorithm integrated switching system, we reduce the original four-dimensional character factors to only one (delay range) by summing up the required PEs of individual delay ranges in the collected dataset. There are 1000 data for each delay range, so we divide the summation by 1000 to obtain the average PE number, which is represented by the y-axis in Fig. \ref{pic:classifier_evaluation}. The x-axis is the delay range. This figure compares the serial paradigm, parallel paradigm, real switching system supported by the trained best classifier, and the ideal switching system. The purple line fitting the real switching system of 91.69\% accuracy is very close to the ideal pink line where data is collected from the label of the dataset. The blue and green lines trend presents the two paradigms' different sensitivity and the advantage area to delay range. By using the classifier switching between them, the switching system automatically takes the strengths of both paradigms and achieves better memory performance than any of the individual paradigms.

Fig. \ref{pic:classifier_evaluation} merely depicts the advantage of mapping one SNN layer with the classifier-integrated switching system. When we apply this approach to the whole neural network, especially a large one with the combination of various layers, the advantage in saving memory and multi-core processors will be greatly increased. For instance, the gesture recognition SNN model with 2048-20-4 structure and 3.16\% weight density mentioned in \cite{ref_jennifer_frontiers} needs 9 PEs on the serial paradigm, 5 PEs on the parallel paradigm, and only 4 PEs by deploying the switching system. 

Improving the spatial performance will provide more possibilities for deploying more complex biological networks and processing multi-tasks simultaneously on the SpiNNaker2 multi-core supercomputer. Our future work will integrate the temporal and energy performances as evaluation criteria into this switching system.


\section{Conclusion} \label{sec_conclusion}
This paper proposes integrating and switching the serial and parallel mapping paradigms of SNN in one system. During the host compilation phase, we deploy a classifier with 91.69\% accuracy to accelerate the switching decision during compiling and deduce the average storage occupation on SpiNNaker2. This fast switching system enables mapping larger-scale SNNs and more tasks on the multi-core supercomputer and provides the methodology of maximizing the benefits of the heterogeneous hardware system for neuromorphic application.


\section*{Acknowledgments}
This work was partially funded by 
the German Research Foundation (DFG, Deutsche Forschungsgemeinschaft) as part of Germany’s Excellence Strategy – EXC 2050/1 – Project ID 390696704 – Cluster of Excellence “Centre for Tactile Internet with Human-in-the-Loop” (CeTI) of Technische Universität Dresden, the EIC Transition under the ”SpiNNode” project (grant number 101112987), and Center for Scalable Data Analytics and Artificial Intelligence (ScaDS.AI). 

\bibliography{references} 

\begin{thebibliography}{10}
\providecommand{\url}[1]{#1}
\csname url@samestyle\endcsname
\providecommand{\newblock}{\relax}
\providecommand{\bibinfo}[2]{#2}
\providecommand{\BIBentrySTDinterwordspacing}{\spaceskip=0pt\relax}
\providecommand{\BIBentryALTinterwordstretchfactor}{4}
\providecommand{\BIBentryALTinterwordspacing}{\spaceskip=\fontdimen2\font plus
\BIBentryALTinterwordstretchfactor\fontdimen3\font minus \fontdimen4\font\relax}
\providecommand{\BIBforeignlanguage}[2]{{%
\expandafter\ifx\csname l@#1\endcsname\relax
\typeout{** WARNING: IEEEtran.bst: No hyphenation pattern has been}%
\typeout{** loaded for the language `#1'. Using the pattern for}%
\typeout{** the default language instead.}%
\else
\language=\csname l@#1\endcsname
\fi
#2}}
\providecommand{\BIBdecl}{\relax}
\BIBdecl

\bibitem{loihi}
M.~Davies \emph{et~al.}, ``Loihi: A neuromorphic manycore processor with on-chip learning,'' \emph{IEEE Micro}, vol.~38, no.~1, pp. 82--99, 2018.

\bibitem{brainscales_ref1}
\BIBentryALTinterwordspacing
C.~Pehle \emph{et~al.}, ``The brainscales-2 accelerated neuromorphic system with hybrid plasticity,'' \emph{Frontiers in Neuroscience}, vol.~16, 2022. [Online]. Available: \url{https://www.frontiersin.org/articles/10.3389/fnins.2022.795876}
\BIBentrySTDinterwordspacing

\bibitem{brainscales_ref2}
S.~Schmitt \emph{et~al.}, ``Neuromorphic hardware in the loop: Training a deep spiking network on the brainscales wafer-scale system,'' in \emph{2017 International Joint Conference on Neural Networks (IJCNN)}, 2017, pp. 2227--2234.

\bibitem{spi1_ref}
S.~B. Furber \emph{et~al.}, ``The spinnaker project,'' \emph{Proceedings of the IEEE}, vol. 102, no.~5, pp. 652--665, 2014.

\bibitem{francisco_1}
F.~Naveros \emph{et~al.}, ``A spiking neural simulator integrating event-driven and time-driven computation schemes using parallel cpu-gpu co-processing: A case study,'' \emph{IEEE Transactions on Neural Networks and Learning Systems}, vol.~26, no.~7, pp. 1567--1574, 2015.

\bibitem{GENN_for_snn}
\BIBentryALTinterwordspacing
J.~C. Knight and T.~Nowotny, ``{GPUs} outperform current hpc and neuromorphic solutions in terms of speed and energy when simulating a highly-connected cortical model,'' \emph{Frontiers in Neuroscience}, vol.~12, 2018. [Online]. Available: \url{https://www.frontiersin.org/articles/10.3389/fnins.2018.00941}
\BIBentrySTDinterwordspacing

\bibitem{ref_jennifer_aicas23}
J.~Huang \emph{et~al.}, ``Efficient algorithms for accelerating spiking neural networks on mac array of spinnaker 2,'' in \emph{2023 IEEE 5th International Conference on Artificial Intelligence Circuits and Systems (AICAS)}, 2023, pp. 1--5.

\bibitem{ref_jennifer_frontiers}
\BIBentryALTinterwordspacing
J.~Huang \emph{et~al.}, ``Efficient snn multi-cores mac array acceleration on spinnaker 2,'' \emph{Frontiers in Neuroscience}, vol.~17, 2023. [Online]. Available: \url{https://www.frontiersin.org/articles/10.3389/fnins.2023.1223262}
\BIBentrySTDinterwordspacing

\bibitem{francisco_2}
\BIBentryALTinterwordspacing
F.~Naveros \emph{et~al.}, ``Event- and time-driven techniques using parallel cpu-gpu co-processing for spiking neural networks,'' \emph{Frontiers in Neuroinformatics}, vol.~11, 2017. [Online]. Available: \url{https://www.frontiersin.org/articles/10.3389/fninf.2017.00007}
\BIBentrySTDinterwordspacing

\bibitem{Mayr_ref}
C.~Mayr, S.~Hoeppner, and S.~Furber, ``{SpiNNaker~2}: A 10 million core processor system for brain simulation and machine learning,'' \emph{arXiv preprint arXiv:1911.02385}, 2019.

\bibitem{Amir}
\BIBentryALTinterwordspacing
A.~Rostami \emph{et~al.}, ``E-prop on spinnaker 2: Exploring online learning in spiking rnns on neuromorphic hardware,'' \emph{Frontiers in Neuroscience}, vol.~16, 2022. [Online]. Available: \url{https://www.frontiersin.org/articles/10.3389/fnins.2022.1018006}
\BIBentrySTDinterwordspacing

\bibitem{Yexin_new_ref}
Y.~Yan \emph{et~al.}, ``Comparing loihi with a spinnaker 2 prototype on low-latency keyword spotting and adaptive robotic control,'' \emph{Neuromorphic Computing and Engineering}, vol.~1, no.~1, p.~16, July 2021.

\bibitem{Zeinolabedin_ref}
S.~M.~A. Zeinolabedin \emph{et~al.}, ``A 16-channel fully configurable neural soc with 1.52 $\mu$w/ch signal acquisition, 2.79 $\mu$w/ch real-time spike classifier, and 1.79 tops/w deep neural network accelerator in 22 nm fdsoi,'' \emph{IEEE Transactions on Biomedical Circuits and Systems}, vol.~16, no.~1, pp. 94--107, 2022.

\bibitem{spynnaker_ref}
O.~Rhodes \emph{et~al.}, ``{sPyNNaker}: A software package for running pynn simulations on spinnaker,'' \emph{Frontiers in Neuroscience}, vol.~12, p. 816, 2018.

\bibitem{lif_model}
G.~Bellec \emph{et~al.}, ``A solution to the learning dilemma for recurrent networks of spiking neurons,'' \emph{Nature communications}, vol.~11, 2020.

\bibitem{microcircuit2018}
\BIBentryALTinterwordspacing
S.~J. van Albada \emph{et~al.}, ``Performance comparison of the digital neuromorphic hardware spinnaker and the neural network simulation software nest for a full-scale cortical microcircuit model,'' \emph{Frontiers in Neuroscience}, vol.~12, 2018. [Online]. Available: \url{https://www.frontiersin.org/articles/10.3389/fnins.2018.00291}
\BIBentrySTDinterwordspacing

\end{thebibliography}
\bibliographystyle{IEEEtran}

\end{document}